\documentclass[prb,nofootinbib,twocolumn,superscriptaddress]{revtex4} 

\usepackage{graphicx}
\usepackage{dcolumn}
\usepackage{bm}
\usepackage{threeparttable}
\usepackage{times}
\usepackage{mathptmx}
\usepackage{lscape}
\usepackage{natbib}
\usepackage{amsmath}
\usepackage{amssymb}
\usepackage{braket}

\def\degree{\kern-.2em\r{}\kern-.3em}

\begin{document}


\title{ Accurate Prediction of Potential Energy Surface via Thermodynamically Equilibrium Structure }

\author{Koretaka Yuge}
\affiliation{
Department of Materials Science and Engineering,  Kyoto University, Sakyo, Kyoto 606-8501, Japan\\
}%

\begin{abstract}
{  In order to predict the potential energy surface (PES) from measured structure in equilibrium state, one should typically perform trial-and-error statistical thermodynamic simulation with assumed multibody interactions. Very recently, we derive  map from a set of equilibrium structure in crystalline solids to that of corresponding PES in explicit matrix form, where the PES can be inversely determined from the measured structure. The practical problem to construct the map appears when system size of measured structure is not sufficiently large, which results in non-trivial treatment of asymmetry problem in the map. The present study proposes alternative approach to avoiding treatment of the asymmetry problem, demonstrating more accurate prediction of the PES than the map constructed by explicitly treating the asymmetry. 
 }
\end{abstract}


\maketitle

\section{Introduction}
Let us consider a classical system, where total energy is the sum of potential energy and kinetic energy.   
In order to quantitatively describe macroscopic structure with structural degree of freedom of $g$, we introduce corresponding complete coordination of $\left\{h_1, \cdots, h_g \right\}$ (orthonormality is not required here.)
In crystalline solids, since the lifetime of a particular configuration is long enough to achieve vibrational equilibrium, canonical average of macroscopic structure along coordination $h_r$, $H_r\left(\beta\right)$, can be given by
\begin{eqnarray}
\label{eq:h}
H_r\left(\beta\right) &=& \sum_d h_r^{\left(d\right)} \exp\left(-\beta E^{\left(d\right)}\right) \nonumber \\
&\simeq&  \sum_s h_r^{\left(s\right)} \exp\left\{-\beta\left(U^{\left(s\right)} + F_{\textrm{vib}}^{\left(s\right)}\left(\beta\right) \right)  \right\}. 
\end{eqnarray}
Here, $\beta$ means $\left(k_{\textrm{B}}T\right)^{-1}$, $d$ and $s$ respectively denotes microscopic state on phase and configuration space, and $F_{\textrm{vib}}^{\left(s\right)}$ represents vibrational free energy on configuration $s$. 
Since the number of possible configuration $s$ astronomically increases with increase of system size, direct determination of $\left\{H_r\left(\beta\right) \right\}$ from first-principles calculation is far from practical, thus a variety of thermodynamic simulation have been performed to efficiently sample dominant microscopic states to determine $H_r\left(\beta\right)$, such as Monte Carlo simulation with Metropolis algorism.\cite{mc1,mc2} 

Basically, to predict macroscopic structure, one first prepares a set of multibody interaction between constituents (or alternatively, prepare PES itself, $\tilde{U}_j$), then perform thermodynamic simulation. This corresponds to obtain partial information of map
\begin{eqnarray}
\Gamma : \mathbf{\tilde{U}}\ni \tilde{U}_j \mapsto \tilde{H}_j \in \mathbf{\tilde{H}}
\end{eqnarray}
restricted to the prepared PES, $\tilde{U}_j$. 
Here, $\mathbf{\tilde{U}}$ and $\mathbf{\tilde{H}}$ respectively denotes a set of PES and a set of macroscopic structure in equilibrium state. 
The problem to straightforwardly predict PES, $\tilde{U}_k$, from measured structure of $\tilde{H}_k$ is that (i) map $\Gamma$ is generally unknown from information about $\tilde{U}_j \mapsto \tilde{H}_j$ for a restricted set of $j$, and (ii) the condition is unclear where the map $\Gamma$ becomes bijective: e.g., when $\tilde{H}_j$ corresponds to ground-state ordered structure, $\Gamma$ is not bijective even for considering low-dimensional configuration space.\cite{cp}
Therefore, in order to practically predict $\tilde{U}_k$ from $\tilde{H}_k$, one should perform trial-and error thermodynamic simulation to find optimal $\tilde{U}_k$ that provides minimum deviation in predicted structure from observed $\tilde{H}_k$. 

Very recently, we propose a theoretical approach enabling to predict PES by directly construct map $\Gamma$ in a matrix form, from a single measured structure in disordered states. 
The significant advantage of the proposed approach is that it does not require trial-and-error or multiple simulations to construct $\Gamma$. 
However, it can be practically difficult to construct accurate map in a matrix form when the provided system size is not sufficiently large where corresponding CDOS with no-interaction slightly deviates from multidimensional gaussian. 
In the present study, we propose alternative approach to accurately construct the map $\Gamma$ in a matrix form, avoiding to explicitly treat this problem. 
Applicability of the proposed approach is demonstrated by comparing the prediction of potential energy by straightforward construction of map $\Gamma$. 


\section{Derivation and Discussions}

Since the measured macroscopic structure for crystalline solids typically corresponds to specify one microscopic structure $s$, information of PES that can be extracted from the measured structure can  
be given by the following expression: 
\begin{eqnarray}
\label{eq:u}
U\left(s, \beta\right) = \sum_r \Braket{F\left(\beta\right)|h_r} h_r^{\left(s\right)},
\end{eqnarray}
where $F\left(\beta\right) = U + F_{\textrm{vib}}\left(\beta\right)$, and $\Braket{\cdot|\cdot}$ denotes inner product, i.e., trace over microscopic states on configuration space with defined metric.
In our previous studies,\cite{lsi,emrs,yuge-RM,gps} we have derived that when configurational DOS (CDOS) for \textit{non-interacting} system is well-characterized by multidimensional gaussian under constant spatial constraint, canonical average of structure along $h_r$ for \textit{interacting} system with the same class of spatial constraint is universally given by
\begin{eqnarray}
\label{eq:emrs}
H_r\left(\beta\right) \simeq \Braket{h_r}_1 \mp \sqrt{\frac{\pi}{2}} \Braket{h_r}_2 \beta \left\{ U_r^{\left(\pm\right)} - \Braket{U}_1 \right\}, 
\end{eqnarray}
where $\Braket{\cdot}_1$ and $\Braket{\cdot}_2$ respectively denotes taking average and standard deviation over microscopic states on configuration space without weight of Boltzmann factor, and 
$U_r^{\left(\pm\right)}$ denotes energy of special microscopic state, which we call "projection state" (PS), which can be exactly given by inner product form:
\begin{eqnarray}
\label{eq:upm}
U_r^{\left(\pm\right)}\left(\beta\right) = \sum_q \Braket{F\left(\beta\right)|h_q} \Braket{h_q}_r^{\left(\pm\right)},
\end{eqnarray}
where $\Braket{\cdot}_r^{\left(+\right)}$ ($\Braket{\cdot}_r^{\left(-\right)}$) denotes taking partial average over microscopic states on configuration space, whose structure satisfies $h_r \ge \Braket{h_r}_1$ 
($h_r \le \Braket{h_r}_1$). 
The reason of $H_r\left(\beta\right)$ given by two ways in Eq.~(\ref{eq:u}) certainly comes from the slight deviation of practical CDOS for non-interacting system under typical spatial constraints (such as fcc, bcc, hcp and diamond lattice) from multidimensional gaussian, resulting in asymmetry of $U_r^{\left(+\right)}\left(\beta\right)$ with respect to $\Braket{U}_1$. 
Note that when CDOS exactly takes multidimensional gaussian, Eq.~(\ref{eq:emrs}) becomes exact in classical systems, and the result obtained from $U_r^{\left(+\right)}$ becomes exactly the same as that from $U_r^{\left(-\right)}$. 
From Eqs.~(\ref{eq:u})-(\ref{eq:upm}), we can immediately obtain the following relationship between PES and measured structure in equilibrium state:
\begin{eqnarray}
\label{eq:inv}
\left\{ \overline{H_1}\left(\beta\right), \cdots, \overline{H_g}\left(\beta\right)  \right\} &=& \left\{ \Braket{F\left(\beta\right)|h_1},\cdots, \Braket{F\left(\beta\right)|h_g} \right\}\cdot \bm{\Gamma}^{\left(\pm\right)} \nonumber \\
\Gamma_{jk}^{\left(\pm\right)} &=& \gamma^{\left(\pm\right)} \Braket{h_k}_2 \left(\Braket{h_j}_k^{\left(\pm\right)} - \Braket{h_j}_1 \right),
\end{eqnarray}
where $\gamma^{\left(\pm\right)}= \mp \beta \sqrt{\left(\pi/2\right)}$, and $\overline{H_r}\left(\beta\right) = H_r\left(\beta\right) - \Braket{h_r}_1$. 
Note that $\bm{\Gamma}$ can be constructed without any information about energy (or interactions), since we can know 
$\Braket{h_k}_2$, $\Braket{h_j}_k^{\left(\pm\right)}$ and $\Braket{h_j}_1$ without any information about energy or temperature. 
This means that when spatial constraint on constituents of the system is once given, we can \textit{a priori} know map from PES to equilibrium structure. 

Although the map $\bm{\Gamma}$ can be in principle numerically constructed using Eq.~(\ref{eq:inv}), practical consideration is that it is non-trivial whether $\Braket{h_j}_k^{\left(+\right)}$ or $\Braket{h_j}_k^{\left(-\right)}$ is used to construct $\bm{\Gamma}^{\left(\pm\right)}$ for a practical finite-size system. When the size of system is sufficiently large, this problem can be neglected since effect of asymmetry naturally disappears, i.e., $\Braket{h_j}_k^{\left(+\right)} - \Braket{h_j}_1 \simeq \Braket{h_j}_1 - \Braket{h_j}_k^{\left(-\right)}$. However, provided measured structure to predict PES does not always have sufficient system size to satisfy the above condition. 
In the present study, we propose alternative approach that can predict PES more accurately than $\bm{\Gamma}$ constructed based on direct estimation of $\Braket{h_j}_k^{\left(+\right)}$ or $\Braket{h_j}_k^{\left(-\right)}$ for intermediate system size. 

Let us consider that the system where the asymmetry $\Braket{h_j}_k^{\left(+\right)} - \Braket{h_j}_1 \neq \Braket{h_j}_1 - \Braket{h_j}_k^{\left(-\right)}$ appears. To avoid this problem, we here employ that $\Gamma$ can be given by another form avoiding to explicitly use $\Braket{h_j}_k^{\left(\pm\right)}$,\cite{spes} namely,
\begin{eqnarray}
\label{eq:cov}
\Gamma_{jk} \simeq -\beta \left(1-e^{-\left(q_k^{\textrm{max}}/\sqrt{2}\Braket{q_k}_2\right)^2}\right) \Xi,
\end{eqnarray}
where $q_k^{\textrm{max}}$ denotes maximum value that $q_k$ can take for non-interacting system, which can also be known \textit{a priori} when spatial constraint is given, and $\Xi$ corresponds to covariance matrix for the CDOS. The approximate equality in Eq.~(\ref{eq:cov}) comes from the differences in integrating one dimensional CDOS along $q_j$, between $\int_{-\infty}^\infty dq_j$ and $\int_{q_j^{\textrm{min}}}^{q_j^{\textrm{max}}}dq_j$, whose deviation converges much faster than the difference between $\Braket{h_j}_k^{\left(+\right)} - \Braket{h_j}_1$ and $\Braket{h_j}_1 - \Braket{h_j}_k^{\left(-\right)}$ in terms of constructing $\Gamma$. We can expect the advantages to employ Eq.~(\ref{eq:cov}) of (i) element of $\bm{\Gamma}$ can be uniquely determined, and (ii) covariance for CDOS is  
typically stable with the changes in periodic boundary condition and system size: It is thus expected that we would obtain more accurate PES from Eq.~(\ref{eq:cov}) than using non-trivial choice of $\Braket{h_j}_k^{\left(+\right)}$ or $\Braket{h_j}_k^{\left(-\right)}$ to construct $\bm{\Gamma}$ when provided system size is not sufficiently large. 

We finally confirm the applicability of using Eq.~(\ref{eq:cov}) to inversely predict PES. We prepare 2-dimensional square lattice with artificially-introduced multibody interactions for three different system size of $N=256$, 512 and 1024. We construct three type of matrix $\bm{\Gamma}$ at around quadruple transition temperature: $\bm{\Gamma}^+$  by using a set of $\Braket{h_j}_k^{\left(+\right)}$, $\bm{\Gamma}^-$ by a set of $\Braket{h_j}_k^{\left(-\right)}$, and $\bm{\Gamma}^0$ by using Eq.~(\ref{eq:cov}). 

\begin{figure}[h]
\begin{center}
\includegraphics[width=0.9\linewidth]
{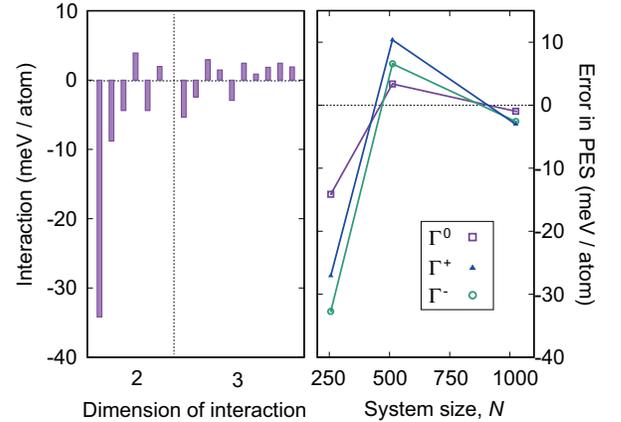}
\caption{Left: Used multibody interactions on square lattice at equiatomic composition. Right: Error in prediction of PES by using three types of map, $\bm{\Gamma}^+$, $\bm{\Gamma}^-$ and $\bm{\Gamma}^0$ as a function of system size, $N$.}
\label{fig:pes}
\end{center}
\end{figure}

Figure~\ref{fig:pes} shows the used multibody interactions (left-hand side), and the accumulating deviation for inversely predicted PES over pairwise interactions by inner product form measured from the original value of their interactions, as a function of system size. It can be clearly seen that while system size is large ($N=1024$), three types of matrix, $\bm{\Gamma}^+$, $\bm{\Gamma}^-$ and $\bm{\Gamma}^0$ provides around the same accuracy of the PES, the error becomes more enhanced for $\bm{\Gamma}^+$ and $\bm{\Gamma}^-$ than for $\bm{\Gamma}^0$ when the system size decreases. These results certainly indicate that when the provided size of measured structure is not sufficiently large (in terms of CDOS satisfying multidimensional gaussian), the proposed approach of using $\bm{\Gamma}^0$ can provide more accurate prediction of PES than directly construct map $\bm{\Gamma}$ by its definition. 

Finally, we propose practical approach to further modify predicting PES when the measured structure is in intermediate state between ordered and well-disordered states. In this case, it is highly expected that precidted PES can deviate from original PES due mainly to the contribution from anharmonicity in structural degree of freedom.\cite{spes} For the microscopic structure in harmonic state, since its deviation from ideal harmonic state is interpreted as perturbed, modification of PES can be effectively performed so that predicted (statistically-averaged) structure multiplied by a certain constant, $g$, can well-reproduce the measure microscopic structure. For instance, Fig.~2 shows the predicted structure based on PES obtained by measured structure at $T=4, 2,1.2$ ($T$ is normalized based on transition temperature) compared with the measured structure. When we simply determine the factor $g$ to minimize the Euclide distance between measured and predicted structure for all coordination considered, the results becomes as shown in Fig.~3. We can clearly see that introducing the factor $g$ can effectively modify predicted structure for low as well as high temperature prediction. This approach can be useful especially when temperature for measured data is restricted to relatively low with respect to the transition temperature, where we can a priori know whether the measured structure is within harmonic region in structural degree of freedom or not without any informatin about temperature or energy.

\begin{figure}[h]
\begin{center}
\includegraphics[width=0.99\linewidth]
{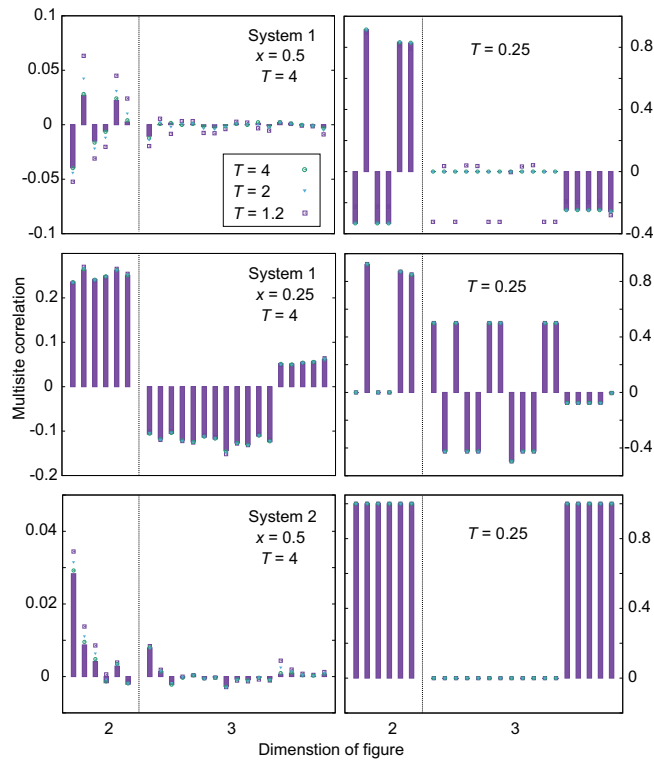}
\caption{Predicted structure (open figures) obtained by PES using microscopic structures measured at different temperature, compared with measured structure on multiple compositions $x$ on two-dimensional triangular lattice. }
\label{fig:2}
\end{center}
\end{figure}

\begin{figure}[h]
\begin{center}
\includegraphics[width=0.99\linewidth]
{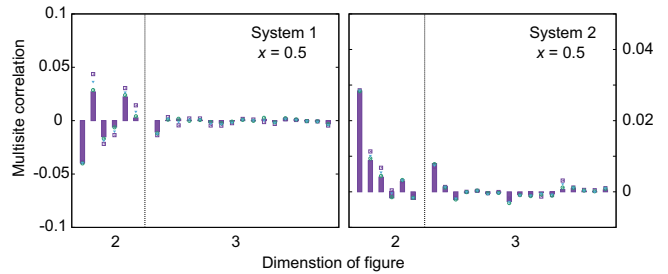}
\caption{Predicted structure modified based on a simple g-factor.}
\label{fig:3}
\end{center}
\end{figure}


\section{Conclusions}
When information about measured structure in equilibrium state is provided, we propose theoretical approach, predicting potential energy surface (PES) by extending our previous approach. 
The proposed approach can avoid non-trivial treatment of asymmetry problem in the map from structure-to-PES transformation found in our previous study. 
Consequently, especially when the provided system size is not sufficiently large where the asymmetry appears, the present approach can provide more accurate prediction of PES than the map directly constructed by explicitly treating the asymmetry.

\section*{Acknowledgement}
This work was supported by a Grant-in-Aid for Scientific Research (16K06704) from the MEXT of Japan, Research Grant from Hitachi Metals$\cdot$Materials Science Foundation, and Advanced Low Carbon Technology Research and Development Program of the Japan Science and Technology Agency (JST).

\end{document}